\documentclass[12pt,epsfig]{article}
\usepackage{graphicx}
\usepackage{amssymb}
\def\beq{\begin{equation}}
\def\eeq{\end{equation}}
\def\bea{\begin{eqnarray}}
\def\eea{\end{eqnarray}}

\renewcommand{\thefootnote}

\begin{document}
\begin{center}
{\Large \bf \sf PT phase transition in a (2+1)-d relativistic system
  }
\end{center}
\begin{center} 
\vspace{1.3cm}
{\sf{B. P. Mandal$^{a}$, \footnote {e-mails address: bhabani.mandal@gmail.com} B. K. Mourya$^{a}$,\footnote{e-mail address: brijeshkumarbhu@gmail.com} K. Ali$^{b}$,\footnote{e-mail address: ali.ksr71@gmail.com} A. Ghatak$^{a}$\footnote{e-mail address: gananya04@gmail.com}}\\
\bigskip
{ $^{a}$ Department of Physics, Banaras Hindu University, Varanasi -221005, India  }\\
{ $^{b}$ Condensed Matter Division, Bhabha Atomic Research Centre, Mumbai 400085, India }\\


\bigskip
\bigskip

\noindent {\bf Abstract}

}
\end{center}

We study a massless Dirac particle with PT symmetric non-Hermitian Rashba interaction in the background of Dirac 
oscillator potential to show the PT phase transition in a (2+1) dimensional relativistic system analytically. PT phase transition occurs when strength of the (i) imaginary Rashba interaction or (ii) transverse magnetic 
field exceed their respective critical values. Small mass gap in the spectrum, consistent with other approaches is generated  as long as the system is in the
unbroken phase. Relativistic Landau levels are constructed explicitly for such a system. 

\medskip
\vspace{1in}

\newpage

\section{Introduction}  
Consistent quantum theory with real energy eigenvalues, unitary time evolution and probabilistic interpretation
for combined Parity (P) and Time reversal (T) symmetric non-Hermitian theories in different Hilbert space equipped 
with positive definite inner product has been the subject of intrinsic research in frontier physics over the last
one and half decades \cite{ben4}-\cite{ben}. The huge success of the complex quantum theory has lead to extension to many other branches 
of physics  and has found many applications \cite{opt1}-\cite{cal}. Such  non-Hermitian PT symmetric systems generally exhibit a phase transition \cite{pt1}-\cite{pt4} ( or more specifically a PT breaking transition ) that separates two parametric regions (i) region of the unbroken PT symmetry in which the entire spectrum is real and eigenstates of the systems respect PT symmetry and (ii) a region of the broken PT symmetry in which the whole spectrum (or a part of it)  appears as complex conjugate pairs and the 
 eigenstates of the systems do not respect PT symmetry. The study of PT phase transition has become extremely 
 important due to the fact that such phase transition and its rich consequences are really observed in variety of
 physical systems \cite{22}-\cite{29}. However most of the analytical studies on PT phase transition are restricted to one dimensional
 non relativistic systems. Few groups \cite{fw1}-\cite{fw2} have studied PT phase transition in higher dimensions for restricted class of 
 non-Hermitian potentials, in non-relativistic situations.
 In the present work we consider a relativistic PT symmetric system in (2+1) dimensions to study the PT phase transition and its consequences.

As a model we consider a relativistic system of a massless Dirac particle in (2+1) dimension \cite{g0}.
Such systems have become extremely important due to the discovery of graphene, a single
layer of the carbon atoms arranged in a honeycomb lattice. The study of graphene has created an avalanche in the frontier
research over the past decade due to its exceptional electronic properties and potentially significant applications \cite{g01}-\cite{g4}.
The low energy excitations of the atoms in the graphene obey massless Dirac equation in (2+1) dimensions. The two 
dimensional graphene sheet is essentially a zero gap semiconductor with Fermi level $E_f$ precisely at $E=0$ with 
a linear dispersion relation given by $E=\hbar v_f k$ where $v_f$ is the Fermi velocity and $k$ is the carrier wave vector.
However a small gap has been found experimentally in the graphene spectrum \cite{mgg}. Such a mass gap is generated by breaking 
the inversion symmetry either (i) by considering spin-orbit interaction (e.g. Rashba interaction) or (ii) by the coupling the system
with quantized circularly polarized fields. In this present work we consider a massless Dirac particle with PT symmetric imaginary Rashba 
interaction \cite{rsbi} in the background of Dirac oscillator (DO) potential \cite{dosc1}-\cite{dosc8} to demonstrate the PT phase transition in (2+1) dimensional 
relativistic system. We analytically show that PT phase transition occurs in such a system when the 
strength of (i) the imaginary Rashba interaction or (ii) the transverse magnetic field exceeds a critical value. However it is challenging to make a realistic experimental setup to observe the PT-phase transition
for the system of graphene with imaginary Rashba interaction.  Small mass gap which is consistent with other approaches are also generated  as long as the system is in PT unbroken phase. The mass gap vanishes at the phase transition point and becomes
imaginary in the broken phase of the system. We construct the solutions for the other valley of the system and show the PT phase transition in the similar fashion. Relativistic lowest order Landau levels are constructed explicitly for the system.
Connection of this system to well-known Jaynes-Cumming (JC) \cite{jc1}-\cite{jc2} model in optics is also indicated. \\
Now we present the plan of the paper. In section (2) we introduce the model and its symmetries. PT phase transition and generation of mass
gap in the spectrum are discussed in section (3). Section (4) is kept for conclusions.

\section{ The Model and PT Symmetry} 
The massless Dirac particle with imaginary Rashba interaction in the transverse static$^1$ \footnote{$^1$ Various solutions of Dirac equation in presence of inhomogeneous\cite{inh1}-\cite{inh3} as well as
time dependent \cite{timed1}-\cite{timed3} magnetic field   have also been obtained for hermitian model} magnetic field in the background of DO 
potential is described by the Hamiltonian  
\beq
H  = v_f[\vec{\sigma} .(\vec {\Pi} - i K_1 \vec{ r}  \beta ) + i \lambda (\vec { \sigma }\times \vec {\Pi }).\hat z ,
\label{one}
\eeq
where $\vec{\Pi} =  \vec {p} + \frac{e \vec{A} }{c}$, and $K_1$, $\lambda$  are real constants and $v_f$ is Fermi velocity.\\
We consider the magnetic field along  z-direction, $ B = {B_0}\hat k$ and choose the vector potential as 
$\vec A =(-\frac{B_0 y}{2} , \frac {B_0 x}{2},0) $ in the symmetric gauge. The term linear in $\vec{r}$ is 
DO potential \cite{dosc1}  as in the non-relativistic limit, it  
reduces to simple harmonic oscillator (SHO) potential with strong spin-orbit coupling. Initially, DO was introduced 
in the context of many body theory, mainly in connection with quark confinement models in QCD  \cite{qcd}. Later the subsequent studies
have revealed several exciting properties connected to the symmetries of the theory and DO has found
many physical applications in various branches of physics \cite{dosc2}-\cite{dosc8}. Non-Hermitian version of  Dirac equation is also considred earlier.In particular, $(2+1)$ dimensional massless Dirac equation in the presence
of complex vector potential and its bound state solutions are discussed in \cite{dosc6}.  Exact solutions for the bound states of a graphene Dirac electron
in various magnetic field with translational symmetry are obtained in  \cite{new1}.  \\ 
The Hamiltonian in component form is written as 
\begin{eqnarray}
H & = & v_f(\sigma_x \Pi _x + \sigma_y \Pi _y)-i K_1 v_f(\sigma_x x + \sigma_y y )\beta + i \lambda (\sigma_x \Pi _y - \sigma_y \Pi _x)
\label{two}
\end{eqnarray}
 where, $\Pi _x =  p_x - \frac{e B_0 y }{2c}$ and \quad $\Pi _y =  p_y + \frac{ e B_0 x }{2c}$.\\
 Now we discuss the symmetry properties of this model.
Parity transformation, 
\begin{eqnarray}
 \left(\begin {array}{clcr}
x' \\
y'\\
\end{array} \right) = A\left(\begin {array}{clcr}
x \\
y\\
\end{array} \right) \nonumber
\end{eqnarray}
is an improper Lorentz transformation (i.e. det$ A =-1 $) and is defined in two alternative ways in two dimension as,
\begin{equation}
 P_1 : x \longrightarrow  -x, \ \ y\longrightarrow  y,  \ \ p_x\longrightarrow - p_x,  \ \ p_y\longrightarrow p_y.
\label{three}
\end{equation} 
\begin{equation}
 P_2 : x\longrightarrow  x, \ \ y\longrightarrow  -y , \ \ p_x\longrightarrow  p_x,  \ \ p_y\longrightarrow -p_y.
\label{four}
\end{equation}
Under such parity transformations Dirac wavefunctions transform as\\
\begin{eqnarray}
P_1 \psi(x,y,t) =\sigma _y \psi(-x,y,t) \nonumber \\
P_2 \psi(x,y,t) =\sigma _x \psi(x,-y,t) 
\end{eqnarray}
such that free Dirac equation remains invariant under $P_1$ and $P_2$. The time reversal transformation 
$(t \longrightarrow  -t, \ \ i\longrightarrow  -i,  \ \ p_x\longrightarrow - p_x,  \ \ p_y\longrightarrow -p_y )$ in 
(2+1) dimension Dirac theory is defined as $T= i\sigma _y  \tilde K $, where $\tilde K$ is complex conjugation operation 
such that, $T \psi(x,y,t) = i \sigma _y \tilde K \psi(x,y,-t) = i \sigma _y  \psi^{\star}(x,y,-t)$. It is straight forward to check that
\begin{eqnarray}
H ^\dagger& = & v_f(\sigma_x \Pi _x + \sigma_y \Pi _y)-i K_1 v_f(\sigma_x x + \sigma_y y )\beta - i \lambda (\sigma_x \Pi _y - \sigma_y \Pi _x)\nonumber \\
 & \not =&  H \nonumber
\end{eqnarray} 
 This system, however is invariant under both $P_1T $ and $P_2T $ as 
\begin{eqnarray}
P_1 T H (P_1 T)^{-1}& = & H \nonumber \\
P_2 T H (P_2 T)^{-1}& = & H \nonumber
\end{eqnarray}
A massless relativistic  system has two degenerate but inequivalent solutions known as valleys of the system. For the present  system in this work
there exists another Dirac equation  corresponding to the other valley. The Hamiltonian corresponding to other valley is 
obtained by taking a time reversal transformation of H \cite{PRL1, PRL2} as,
\begin{equation}
\tilde{H}   =THT^{-1}= v_f(\sigma_x \tilde{\Pi} _x  + \sigma_y \tilde{\Pi} _y)-i K_1 v_f(\sigma_x x + \sigma_y y )\beta - i \lambda (\sigma_x \tilde{\Pi} _y- \sigma_y \tilde{\Pi} _x)
\label{two2}
\end{equation}
where $\tilde{\Pi} _x =  p_x + \frac{e B_0 y }{2c}$ and  $\tilde{\Pi} _y = p_y - \frac{ e B_0 x }{2c}$.
It is straight forward to show that $\tilde {H}\not =\tilde {H}^\dagger$ and $[\tilde {H},P_1 T]=[\tilde {H},P_2 T]=0$.
Now in the next section we find the solution of Dirac equation for both the valleys to show the PT phase transition and  mass gap generation in such systems.

\section{ PT phase transition and generation of mass gap}
In this section we solve the Dirac equation with Hamiltonian H and $\tilde H$ corresponding to the valleys for the
massless particle explicitly
to demonstrate PT phase transition in (2+1) dimensional relativistic system. In both cases we found that mass gap is generated 
due to imaginary Rashba interaction in the background of DO potential as long as the system is in the unbroken phase. We further show
how the different solutions of Dirac equation for both the valleys in this particular model are interrelated. We begin with writing
the Hamiltonians in Eqs.(\ref{two}) and (\ref{two2}) in a compact form on a complex plane $z = (x + i y ) $ as,
\begin{equation}
H  =\left( \begin{array}{clcr}
0 \  \ A\Pi  _z + i C_1 \bar z  \\
B \Pi_{ \bar z}+ iC_2 z  \   \ 0 
\end{array} \right) \ \ , \ \ \tilde{H} = \left( \begin{array}{clcr}
0 \  \ \ \ B\Pi  _z + i C_2 \bar z  \\ 
\\
A \Pi_{ \bar z}+ iC_1 z  \  \ \ \ 0 
\end{array} \right) 
\label{five}
\end{equation} 
where, $A = 2 (v_f-\lambda )$,\quad $B = 2 (v_f+ \lambda )$,\quad $C_1 = K_1 v_f-(v_f-\lambda )\frac{ B_0 e }{2c}$ and\\ 
$C_2 = -K_1 v_f +(v_f+\lambda )\frac{ B_0 e }{2c} $ are constants and canonical conjugate momenta in complex plane $\Pi_ z $ and
$\Pi_{ \bar z}$ are defined as 
$\Pi_ z = -i \hbar  \frac{d} {dz}= \frac {1} {2} (p_x- i p_y)$,\quad $\Pi_{ \bar z} =  -i \hbar  \frac{d} {d \bar z}= \frac {1} {2} (p_x + i p_y)$ with the properties
\begin{equation}
 \left[\bar z,\Pi_ {\bar z}\right] =  i \hbar ; \quad \left[z,\Pi_{ \bar z} \right] = 0 ;\quad 
\left[\bar z,\Pi_{ z}\right] = 0 ; \quad  \left[\Pi_{ \bar z},\Pi_{ z}\right] = 0; \quad \left[z,\Pi_{z} \right] = i \hbar  \quad \quad \label {six}
\end{equation}


\subsection {Solution for the system H}
Now we first present the solutions corresponding to the Hamiltonian $H$. To find the solution of the
Dirac equation corresponding to $H$ we assume a solution of the corresponding Dirac equation $H \psi = E \psi $ in the two components form as,
\begin{equation}
\psi   =\left( \begin{array}{clcr}
\phi    \\
 i \chi \\ 
\end{array} \right) 
\label{seven}
\end{equation}
Dirac equation is then written in components form as 
\begin{equation}
(A\Pi  _z + i C_1 \bar z) (B \Pi_{ \bar z}+ iC_2 z) \phi = E^2 \phi \
\label{eight}
\end{equation} 
\begin{equation}
(B \Pi_{ \bar z}+ iC_2 z) (A\Pi  _z + i C_1 \bar z) \chi= E^2 \chi
\label{nine}
\end{equation} 
Now we look for a solution of the kind
\begin{eqnarray}
\phi & =&  \xi (z,\bar z) e^{d_1 z \bar z} \nonumber  \\
\chi & =&  \eta  (z,\bar z) e^{d_1 z \bar z} 
\label{nine1}
\end{eqnarray}
Substituting these in Eq.(\ref{eight}) or in Eq.(\ref{nine}) and comparing the coefficient of  $(z \bar z)$  in both sides we obtain
$ d_1= \frac {C_1}{A \hbar}$ or $\frac {C_2}{B\hbar}$. For $ d_1= \frac {C_1}{A \hbar}$ the Eqs.(\ref{eight}) and (\ref{nine}) reduce to,
\begin{equation}
-AB\hbar^2  \frac{\partial^2 \xi}{\partial z \partial \bar z}+ K z \frac{\partial \xi}{\partial z}  - (E^2-K) \xi = 0
\label{tan}
\end{equation} 
\begin{equation}
-AB\hbar^2  \frac{\partial^2 \eta }{\partial z \partial \bar z}+ K z \frac{\partial \eta }{\partial z}  - E^2 \eta  = 0
\label{eleven}
\end{equation} 
Similarly for $d_1=\frac {C_2}{B\hbar}$ the Eqs.(\ref{eight}) and (\ref{nine}) reduce to,
\begin{equation}
-AB\hbar^2  \frac{\partial^2 \xi}{\partial z \partial \bar z}- K \bar z\frac{\partial \xi}{\partial \bar z}  - E^2 \xi = 0
\label{twelve}
\end{equation} 
\begin{equation}
-AB\hbar^2  \frac{\partial^2 \eta }{\partial z \partial \bar z}- K \bar z \frac{\partial \eta }{\partial \bar  z}  -( E^2 +K)\eta  = 0
\label{thirteen}
\end{equation} 
We find the solution of the Eqs.(\ref{tan}) and (\ref{eleven})(i.e. for $d_1=\frac {C_1}{A\hbar}$ ) as 
\begin{eqnarray}
\xi _n &= &a^{I}_n z^n    \nonumber  \\
\eta  _n &= & a^{I}_n z^{n+1}\qquad \mbox {with}, \quad  E_n^2= ( n+1) (AC_2-BC_1) \nonumber
\end{eqnarray}
where $a^{I}_n$ are real constants. Thus the general solution for the $n^{th}$ level for the case $ d_1= \frac {C_1}{A \hbar}$ is
\begin{equation}
\psi^{I} _n (z,\bar z ,t)=\left( \begin{array}{clcr}
\xi_n     \\
 i \eta_n \\ 
\end{array} \right) e^{\frac {C_1}{A \hbar} {z \bar z}} e^{-\frac {i }{ \hbar} E_n t} =a^{I}_n \left( \begin{array}{clcr}
 z^n     \\
 i  z^{n+1} \\ 
\end{array} \right) e^{\frac {C_1}{A \hbar} {z \bar z}} e^{-\frac {i }{ \hbar} E_n t} 
\label{fou}
\end{equation}
In exactly similar fashion we obtain the solution corresponding to $ d_1= \frac {C_2}{B \hbar}$ as,
\begin{equation}
\psi^{II} _n (z,\bar z ,t)=a^{II}_n \left( \begin{array}{clcr}
\bar z^{n+1}      \\
 i \bar z^{n} \\ 
\end{array} \right) e^{\frac {C_2}{B \hbar} {z \bar z}} e^{-\frac {i }{ \hbar} \tilde{E} _n t} , \ \ \mbox {with}, \quad  \tilde{E}_n^2= ( n+1) (BC_1-AC_2)=-E_n^2
\label{fou2}
\end{equation}

\subsection{ Solution corresponding to other valley}
 
To find the solution corresponding to the Dirac equation for $\tilde H$  we observe that, 
{\large \[H \mathrel{\mathop{--- \longrightarrow }^{\mathrm{A\leftrightarrow B}}_{\mathrm{C_1\leftrightarrow C_2}}} \tilde{H}\]}
Interchanging $A \leftrightarrow B, C_{1}\leftrightarrow C_{2}$ is equivalent to $\lambda \rightarrow -\lambda, K_{1}\rightarrow -K_{1}$ and 
$B_{0}\rightarrow -B_{0}$. Thus it is straight forward to obtain the solutions for the Dirac equation corresponding to $\tilde H$ which describes their valley, by 
using these changes of parameters in the solutions corresponding to Dirac equation with Hamiltonian H.
The solutions for the systems $\tilde H$ are given as, \\
\beq
\mbox{For} \ \ \tilde{d}_1= \frac {C_1}{A \hbar} \ , \ \ \ \tilde{\psi}^{I} _n (z,\bar z ,t)=\tilde{a}^{I}_n \left( \begin{array}{clcr}
\bar z^{n+1}      \\
 i \bar z^{n} \\ 
\end{array} \right) e^{\frac {C_1}{A \hbar} {z \bar z}} e^{-\frac {i }{ \hbar} {E}_n t}
\eeq
\beq
\mbox{For} \ \ \tilde{d}_1= \frac {C_2}{B \hbar} \ , \ \ \ \tilde{\psi}^{II} _n (z,\bar z ,t)=\tilde{a}^{II}_n \left( \begin{array}{clcr}
 z^n     \\
 i  z^{n+1} \\ 
\end{array} \right) e^{\frac {C_2}{B \hbar} {z \bar z}} e^{-\frac {i }{ \hbar} {\tilde {E}}_n t}
\eeq
where ${E}_n^{2} = ( n+1) (AC_2-BC_1)=-{\tilde {E}}_n^{2} $. 
We will be using these results to discuss PT phase transition and generation of mass gap in the next subsection.

\subsection{ PT phase transition}
The  energy eigenvalues for the solutions corresponding  to $ d_1= \tilde d_1= \frac {C_1}{A \hbar} $ in both the valley are,
\begin{equation}
E_n=\pm \sqrt {(n+1)\left [ 2 (v^2_f-\lambda ^2)\frac{B_0 e \hbar}{c}- 4 K_1  v^2_f \hbar \right ]}
\label{fif}
\end{equation}
This indicates a mass gap 
\begin{equation}
\Delta_{0}= \sqrt {2 (v^2_f-\lambda ^2)\frac{B_0 e \hbar}{c}- 4 K_1  v^2_f \hbar} 
\label{mgap}
\end{equation}
is generated between the positive and negative energy solutions due to the interactions present in this theory.
Now we proceed to show that this system passes through a PT phase transition depending on the different parametric values. 
The energy eigenvalues are real for these solutions when
\begin{eqnarray}
\mbox{(i)}\ \lambda ^2 \leq  v_f^2 ( 1-\frac{2 K_1 c}{B_0 e})\equiv\lambda_{c} ^2 \ \ \ \ \mbox{OR} \nonumber \\ 
\mbox{(ii)}\ B_0 >  2 K_1 \frac{ v^2_f  c} {(v^2_f -\lambda ^2) e}\equiv B_{0}^c 
\label{ntwn}
\end{eqnarray}
As long as $E_n$ is real $\psi^{I} _n$ and   $\tilde \psi^{I}_n $ are an eigenstate of both $P_1T$ and $P_2T $
\begin{eqnarray}
P_1T \psi ^I _n (z,\bar z ,t)&= &\sigma_y i \sigma_y \left( \begin{array}{clcr}
\xi_n(-z,-\bar z)    \\
 -i \eta_n(-z,-\bar z) \\ 
\end{array} \right) e^{\frac {C_1}{A \hbar} {z \bar z}} e^{-\frac {i }{ \hbar} E_n t},\quad \mbox {as }P_1T z= -z\nonumber \\
 &= & i (-1)^n \left( \begin{array}{clcr}
\xi_n    \\
 i \eta_n \\ 
\end{array} \right) e^{\frac {C_1}{A \hbar} {z \bar z}} e^{-\frac {i }{ \hbar} E_n t} = i (-1)^n\psi ^I_n (z,\bar z ,t)
\label{sixt}
\end{eqnarray} 
Similarly,
\begin{eqnarray}
P_2T \psi ^I _n (z,\bar z ,t)&= &\sigma_x i \sigma_y \left( \begin{array}{clcr}
\xi_n(z,\bar z)    \\
- i \eta_n (z,\bar z) \\ 
\end{array} \right) e^{\frac {C_1}{A \hbar} {z \bar z}} e^{-\frac {i }{ \hbar} E_n t},\quad\mbox {as }P_2T z= z \nonumber \\
&= & -i \sigma_z \left( \begin{array}{clcr}
\xi_n    \\
 -i \eta_n \\ 
\end{array} \right) e^{\frac {C_1}{A \hbar} {z \bar z}} e^{-\frac {i }{ \hbar} E_n t} \nonumber \\
&= & -  \left( \begin{array}{clcr}
\xi_n    \\
 i \eta_n \\ 
\end{array} \right) e^{\frac {C_1}{A \hbar} {z \bar z}} e^{-\frac {i }{ \hbar}E_n t} = - \psi ^I _n (z,\bar z ,t) 
\label{sixt2}
\end{eqnarray}
Similarly we obtained for the other valley as,
\bea
P_1T \tilde{\psi }^I_n (z,\bar z ,t)&= &-i (-1)^n\tilde{\psi }^I _n (z,\bar z ,t) \nonumber \\
P_2T \tilde{\psi }^I_n (z,\bar z ,t)&= & - \tilde{\psi }^I  _n (z,\bar z ,t)
\eea
Thus for $ d_1= \tilde d_1= \frac {C_1}{A \hbar} $ system corresponding to both the valleys are in unbroken PT phase as long as the coupling for Rashba interaction is
less or equal to $\lambda_c$  or strength strength of the magnetic field is greater or equal to a critical value $ B^{c}_0$. 
Now we observe in this situation the other two solutions (${\psi }^{II}_n, \tilde{\psi }^{II}_n$) (i.e for $ d_1= \tilde d_1= \frac {C_2}{B \hbar} $) correspond to the broken PT phase 
as the eigenvalues for both the valleys are imaginary, $\tilde{E}_n=\pm iE_n$ with real $E_n$. 
In this case $\psi^{II}_n$ and $\tilde \psi^{II}_n$ are not eigenstates of either $P_1T$ or $P_2T$ i.e., 
\beq 
P_i T\psi_n^{II}\not=a_i\psi_n^{II}, \ \ \ P_i T\tilde{\psi}_n^{II}\not=b_i\tilde{\psi}_n^{II} \
\ \mbox{for} \ \ \ i=1,2 
\eeq 
as energy eigenvalue $\tilde E_n$ is imaginary. Hence PT symmetry is broken spontaneously for $d_1=\tilde{d}_1=\frac{C_2}{B\bar h}$.
Thus the system with $d_1=\tilde d_1= \frac {C_1}{A \hbar} $ passes through $P_1T$ and $P_2T$ phase transition where either 
{\bf (a)} strength of the imaginary Rashba interaction is equal to a critical value ($\lambda _c$) or 
{\bf (b)} the external magnetic field $B_{0}$ is equal to a critical value ($B_{0}^c$).

On the other hand if $\lambda > \lambda_c$, for fixed $B_0$ or $ B_0 < B^{c}_0$ for fixed $\lambda $, the solutions for both 
the valleys for $d_1=\tilde d_1= \frac {C_2}{B \hbar} $, i.e. ${\psi }^{II}_n, \tilde{\psi }^{II}_n$ are in the unbroken phase and the solutions corresponding to $d_1=\tilde d_1= \frac {C_1}{A \hbar} $,
i.e. ${\psi }^I_n, \tilde{\psi }^I_n$ are in broken phase.
PT phase transition for $ d_1= \tilde d_1= \frac {C_1}{A \hbar} $ is demonstrated graphically in Fig. 1 by plotting $E_n\  \mbox{vs}\  \lambda$ for fixed value of $B_0$ in Fig. 1(a) and by plotting $E_n\  \mbox{vs} \   B_{0}$ for fixed $ \lambda $ in Fig 1(b). \\

\includegraphics[scale=0.93]{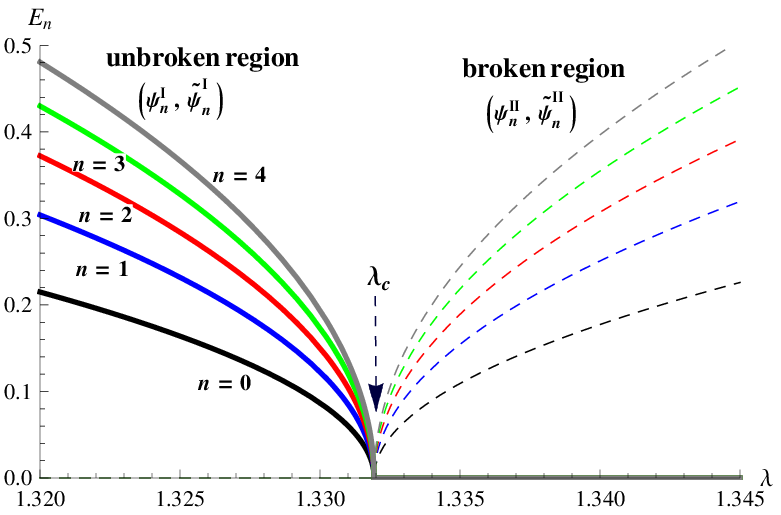} \ \includegraphics[scale=0.93]{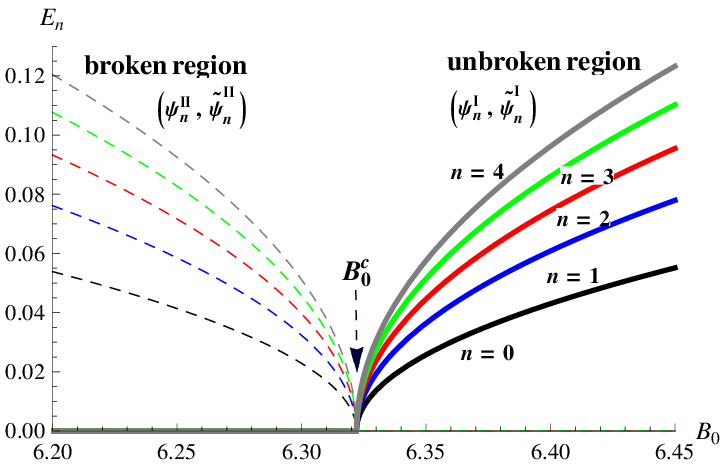}   

\hspace{1.2in} (a) \hspace{2.7in} (b) \\

{\it Fig. 1: PT phase transition for the case for $ d_1= \tilde d_1= \frac {C_1}{A \hbar} $. Fig (a) Real (Solid lines) and imaginary (broken lines) parts of $E_n $ have been plotted with Rashba coupling strength for fixed values of  $B_0 = 100$, $K_1=0.02$, $v_f = 0.01 c$, $e = 1$, $\hbar = 1$, $c = 137$ for $n = 0,1,2,3,4$. Phase transition occurs at $\lambda_c =1.33193 $. Fig (b) Similarly $E_n $ is plotted with  strength of magnetic field $(B_0)$ for fixed values of  $\lambda=0.5$, keeping other parameters same as in Fig 1 (a) for $n = 0,1,2,3,4$. Phase transition occurs at $B_0^{c} = 6.32209$. }\\

Now we would like to concentrate on mass gap $\Delta_{0}$ when the system passes a PT phase transition;

(i) For $d_1=\tilde d_1= \frac {C_1}{A \hbar} $, when $\lambda < \lambda_c$ or $B_0 > B^{c} _0 $ we see from Eq.(\ref{mgap}) mass gap is 
always positive and becomes imaginary when $\lambda > \lambda_{c}$ or $B_0 < B_{0}^c$. Thus a mass gap is generated as long as system 
is unbroken PT phase and given by the expression in in Eq. (\ref{mgap}). Further in unbroken phase $\Delta_{0}$ varies as $\sqrt {B_0}$ 
for large enough magnetic field. This result is consistent with other approaches.

(ii) For $d_1=\tilde d_1= \frac {C_2}{B \hbar} $ when $\lambda > \lambda_c$ or $B_0 < B^{c} _0 $ the system is in unbroken phase  and 
again real mass gap is generated which is consistent with other approaches. Thus mass gap is always generated in this theory 
which is consist with other approaches. In Fig. 2 variation of mass gap is shown with respected to the Rashba coupling strength as well as strength of the magnetic field. Real mass gap for the massless systems is generated as long as the system is in PT unbroken phase. \\

\vspace{0.7in}

\includegraphics[scale=0.90]{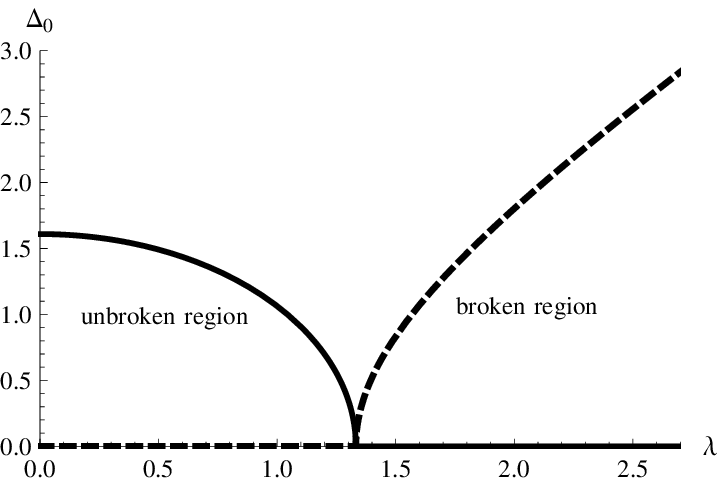} \ \ \ \includegraphics[scale=0.90]{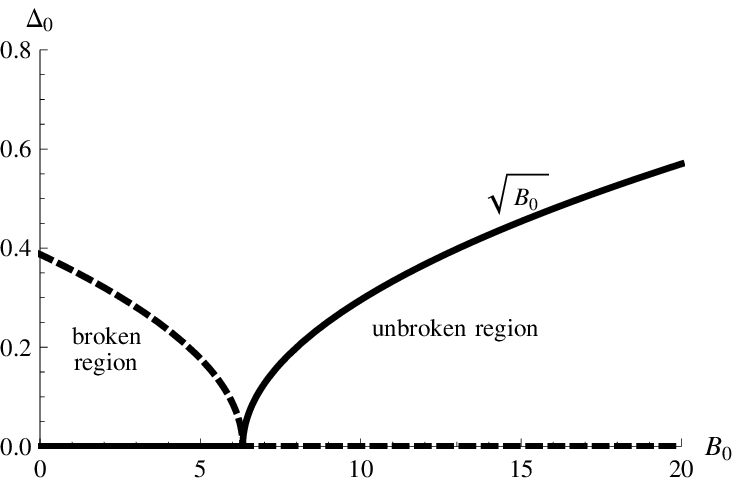}  \\

\hspace{1.1in} (a) \hspace{2.4in} (b) \\

{\it Fig. 2: Variation of mass gap with $\lambda $ and $B_0$ for the case $ d_1= \tilde d_1= \frac {C_1}{A \hbar} $. Fig. (a) Real (Solid line) and imaginary (broken line) part of mass gap is plotted with strength of the Rashba interaction for
fixed values of  $B_0 =100$, $K_1=0.02$, $v_f= 0.01 c$, e=1, $\hbar =1$, $\lambda\leq \lambda_c = 1.33193 $, mass gap is real and system is unbroken PT phase. Fig. (b) Real (Solid line) and imaginary (broken line) part of mass gap is plotted with the strength of magnetic field by keeping the strength of Rashba interaction fixed for a value of  $\lambda=0.5$, $K_1=0.02$, $v_f= 0.01 c$, e=1, $\hbar =1$. $B_{0} \geq B_0^{c} = 6.32209 $, system is in unbroken phase, and real mass gap varies as $\Delta_{0}\propto \sqrt {B_0}$.}\\

\vspace{0.2cm}
{\bf Special cases} : \\ 

(i) In absence of DO $ (K_1 = 0)$, system passes to broken phase when strength of Rashba interaction exceeds Fermi velocity
$\lambda ^2 > v_f^2$. The mass gap in that case (unbroken PT phase) is given by $\Delta_{0}= \sqrt {2 (v^2_f-\lambda ^2)\frac{B_0 e \hbar}{c} }$ which is real as $\lambda ^2 < v_f^2$ for the unbroken phase. Mass gap vanishes at the transition point $\lambda ^2 = v_f^2$.\\
(ii) In absence of Rashba interaction $(\lambda=0)$, the system is Hermitian, but can have real energy eigenvalues only when the magnetic field is sufficiently large i.e. $B_0 >  \frac{  2 K_1 c} {e}$, the mass gap generated in this case is $\Delta_{0}= \sqrt {2 v^2_f \hbar (\frac{B_0 e }{c}- 2 K_{1})}$ which is always real for $B_0 >  \frac{  2 K_1 c} {e}$. In both the cases for large enough magnetic field $\Delta_{0} \propto  \sqrt {B_0}$, which is consistent with other approaches. Again in this case, mass gap vanishes at the transition point, $B_0 = \frac {2 K_1 c} {e}$.  This critical value of magnetic field is same as obtained in \cite{dosc7}. The factor of 2 is due to the choice of vector potential in the symmetric gauge. This discussion in valid for both the valleys. \\

\newpage
{\bf Lowest Landau Levels (LLL)} : \\

The LLLs are obtained by putting the condition $Q_1\chi_0 = 0$, which implies
\begin{equation}
(A\Pi  _z + i C_1 \bar z) \chi_0 = 0.
\label{nitn}
\end{equation} 
We substitute $ \chi_{0} = u_0( z, \bar z) e^{\frac {C_1} {A \hbar} z\bar z} $ to solve Eq.(\ref{nitn}) 
and obtain 
\begin{equation}
\frac{\partial u_0(z,\bar z)}{\partial z } =0,  
\label{20}
\end{equation}
as the defining rule for LLL in this system. We obtain the LLL in the coordinate representation, which is infinitely 
degenerate as
\begin{equation}
\chi_{0}(z, \bar z) = \bar z^l e^{\frac {C_1} {A \hbar} z\bar z} ,\quad l= 0,1,2,.....\infty \\
\label{lll}
\end{equation}
The monomials $\bar z^l$ with $l= 0,1,2,...\infty $  serve as a linearly independent basis. The first and higher excited states
in the coordinate representation are then obtained by applying $Q_2^\dagger$ given in Eq.(\ref{lll}) to the LLL repeatedly. The LLL corresponding to the other valley is obtained by putting 
$(B\Pi _z+iC_2 \bar z)\tilde{\chi_0}=0$. Following similar calculation we obtain LLL $ \tilde{\chi_{0}} 
(z,\bar z)= \bar z^l e^{\frac {C_2} {B \hbar} z\bar z} $,  $l= 0,1,2,...\infty $ and other higher
excited states for this valley.  \\

\vspace{.2cm}
{\bf Connection with JC Model} : \\

We would like to point out the connection of this model with generalized version of well known JC  model for massless particles in optics. The Hamiltonian corresponding to the simple version of (JC)  model is given as 
\begin{equation} 
H_{JC} = g (\sigma _+ a + \sigma _- a^\dagger) + \sigma _z mc^2,
\label{21}
\end{equation}
where $\sigma _{\pm} =  \frac {1}{\sqrt{2}} (\sigma _x \pm i \sigma _y)$ are usual spin raising and lowering operators.
This model is widely used in optics to study the atomic transition in two level systems.\\
Now the Hamiltonian $H$ in Eq.(\ref{five})can be expressed as,
\begin{equation}
H =  \sqrt{K} \left( \begin{array}{clcr}
0 \  \ Q_1  \\
Q_2^\dagger  \   \ 0 
\end{array} \right)
\end{equation}
where, $Q_1= \frac{1}{\sqrt {K}} ( A\Pi  _z + i C_1 \bar z)$ and $Q_2^\dagger= \frac{1}{\sqrt {K}} (B\Pi  _{\bar z} + i C_2  z)$ with $K = ( AC_2- B C_1 ) \hbar$. It is straight forward to check,  
$\left[Q_1,Q_2^{\dagger} \right] = 1$. Note $Q_1 = Q_2$ for real Rashba interaction. The Hamiltonian is not Hermitian but invariant under $P_1 T$ and $P_2 T$ both and is further written as,
\begin{equation} 
H =  \sqrt {K} (\sigma _+ Q_1 + \sigma _- Q_2^\dagger) 
\label{22}
\end{equation}
This is the extended version of JC model for particle with rest mass zero. It is very exciting to note how two completely different
theories are related in this fashion. This provides an alternative approach to study any massless Dirac particle
(such as graphene) with Rashba interaction in the background
 of DO potential in the transverse magnetic field using JC model in quantum optics. For a real Rashba interaction,
$Q_1 = Q_2$ and the extended JC model in Eq.(\ref {22}) exactly coincide with JC model for a massless particle. Similar discussion is also valid for the other valley.

\section {Conclusions} 
We have studied PT phase transition in a (2+1) dimensional relativistic system to enlight different aspects of this important phenomenon.
For this purpose we have considered the massless Dirac particle with imaginary Rashba interaction in the background of 
DO potential in (2+1) dimension. Such a system is non-Hermitian due to the imaginary coupling. We have constructed two different Parity- Time reversal transformations ($P_1T$ and $P_2T$ ) in (2+1) dimensional relativistic system such that the system is invariant under both $P_1T$ and $P_2T$. We have constructed different possible solutions corresponding to both the valleys for this system. The energy eigenvalues are real as long as the eigenstates  
respect $P_1T$ and $P_2T$ symmetry. We further investigate $P_1T$ and $P_2T$ phase transitions in this system to show that system passes from unbroken $P_1T$ and $P_2T$ phases to broken phases depending on the strength of the 
Rashba interaction or transverse magnetic field. The inter-relation of different solutions for both the valleys with respect to PT-phase transition is presented. Even though graphene is zero gap semiconductor with linear dispersion relation, a small mass gap in its spectrum is observed experimentally. In our formulation small mass gap consistent with other approaches is generated for the system of massless Dirac particle when the system is in unbroken phase. The mass gap vanishes at the point of PT phase transition. We would like to point out that results on PT-phase transition presented in this paper
is valid for any system with massless Dirac particle in presence of imaginary Rashba interaction
in the background of Dirac oscillator. In particular our model has some relevance to the possible physics in graphene when the solution for both the valleys are combined.
The lowest order Landau levels for this system have been calculated explicitly to show the infinite degeneracy in such systems. The connection of such system 
with JC model has also been discussed. It will be exciting to correlate other exciting properties of graphene with PT phase transition. \\

{\bf Acknowledgments}
BKM and BPM acknowledge the financial support from the Department of Science and Technology (DST), Govt. of India under SERC project sanction grant No. SR/S2/HEP-0009/2012. AG acknowledges the Council of Scientific \& Industrial Research (CSIR), India for Senior Research Fellowship.

\end{document}